\documentclass[12pt]{iopart}
\usepackage{graphicx}

\begin{document}

\title[The trajectory of light ray under Kerr-Taub-NUT space time.]{The trajectory of light ray under Kerr-Taub-NUT space time.}

\author{Sarani Chakraborty$^1$ \& A. K. Sen$^2$}

\address{$^1$, $^2$ Department of Physics, Assam University, Silchar-788011, Assam, India}
\ead{$^1$sarani.chakraborty.phy@gmail.com; $^2$ asokesen@yahoo.com}
\vspace{10pt}
\begin{indented}
\item[]{\today}
\end{indented}

\begin{abstract}
According to General Relativity, there are three factors namely mass, rotation and charge that
can influence the path of light ray. Many authors showed that there is another factor which can influence the
path of light ray namely gravitomagnetism. Here we discuss the effect of a rotating body with non-zero
(Kerr-Taub-NUT) magnetic field on the motion of light ray. We use the null geodesic of photon method
and obtain the deflection angle of light ray for such a body up to fourth order term in the equatorial plane. Our calculation shows that magnetism has a noticeable effect on the path of light ray. If we set the magnetism equal to zero, our expression of bending angle reduces to the Kerr bending angle. However, we get non-zero bending angle for a hypothetical mass less, magnetic body.
\end{abstract}

%
%
%
%
%

\section{Introduction}

One of the most important predictions of general relativity is deflection of light ray in presence of gravitational mass. Three factors that effect the trajectory of light ray are gravitational mass, rotation and charge. The first order contribution of mass on the path of light ray was calculated by Einstein himself. After Einstein, higher order contribution of mass towards light deflection angle was calculated by Keeton and Petters [1]. Virbhadra and Ellis obtained the lens equation for Schwarzschild black hole [2] and naked singularity [3]. They also calculated the light deflection angle.  Recently, Iyer and Petters [4] calculated it for strong field and found that under weak field approximation their expression matches with that of Keeton and Petters [1].
\\ The light deflection angle for a rotating mass (Kerr mass) was calculated by Iyer and Hansen [5, 6]. Later, Bozza [7] obtained the lensing formula and calculated all other components related to lensing. Azzami et. al [8, 9] calculated the two individual components (parallel and perpendicular to equatorial plane) of light deflection angle in quasi-equatorial regime. Dubey and Sen [10, 11] have used Kerr and Kerr-Newman mass, to show how gravitational redshifts are affected as photon is emitted from various latitudes. Chakraborty and Sen [12] have recently obtained the light deflection angle for a charged, rotating mass in the equatorial plane and showed how deflection angle changes with charge. Chakraborty and Sen [13] recently reported the off equatorial light deflection angle for Kerr mass.  Hasse and Pelrick [14] worked on the lensing by Kerr-Newman mass (charged, rotating) using Morse theory and showed that infinite number of images formed by such mass. Eiroa et al [15] worked on Reissner-Nordstr$\ddot{o}$m (charged, static) mass and calculated light deflection angle under both strong and weak deflection limit. Virbhadra et al [16] worked on Janis-Newman-Winicour (JNW) mass which is a charged, static mass and calculated the light deflection angle up to second order.
\\On the other hand some authors have used material medium approach, where the gravitational effect on light ray was calculated by assuming some effective refractive index assigned to the medium through which light is propagating. With this approach Atkinson [17] studied the trajectory of light ray near a very massive, static and spherically symmetric star. Fischback and Freeman [18] calculated the second order contribution to gravitational deflection by a static mass using the same method. Sen [19] used this method to calculate the gravitational deflection of light without any weak field approximation for Schwarzschild mass. Similar method was used in earlier past by Balaz [20] to calculate the change in the direction of polarization vector of electromagnetic wave passing close to a rotating body. Recently Roy and Sen [21] calculated the trajectory of a light ray in Kerr field using the same material medium approach.
\\All the above mentioned works were done by considering three factors (mass, electric charge and rotation), but various authors showed that not only these three factors but also magnetism can influence space time curvature i.e. path of light ray. In the year of 1963, Newman, Tamburino and Unti [22] first introduced the concept of "generalized Schwarzschild metric". This metric contains one arbitrary parameter in addition to the mass generally known as NUT factor or gravitomagnetic mass. In the same year Misner [23] studied the "generalized Schwarzschild metric" and called it as NUT (named after Newman, Tamburino and Unti) space time. According to him this line element has a Schwarzschild-like singularity, but this singularity is not observed in the curvature tensor [23]. The presence of cross term "$dtd\varphi$" shows, that this space has a strength of gravitomagnetic monopole [24]. Lensing effect of this type of mass was studied by Nouri-Zonoz and Lynden-Bell [25]. They showed that the presence of NUT factor change the observed shape, size and orientation of a image, but they did not show the effect of such body on light deflection angle.
\\On the other hand Kerr-Taub-NUT (KTN) line element [26, 27] represents the solution of Einstein-Maxwell's field equation for a rotating body with non zero magnetic mass. This line element contains three parameters namely mass, rotation parameter and NUT factor $n$. Wei et al. [28] studied numerically the quasi-equatorial lensing by the stationary, axially-symmetric black hole in KTN space time in the strong field limit. Abdujabbarov et al. [29] studied the electromagnetic fields in the KTN space time and in the surrounding space time of slowly rotating magnetized NUT star and obtained analytical solutions of Maxwell equations. Chakraborty and Majumdar [30] derived the exact Lense–Thirring precession frequencies for Kerr, KTN and Taub-NUT space times.
\\In this present work, we studied the KTN line element and obtained the equatorial $(\vartheta=\frac{\pi}{2})$ light deflection angle for such space time geometry up to fourth order term which is a function of mass, rotation parameter and the NUT factor. We also studied the variation of the light deflection angle as a function of NUT factor. For zero NUT factor our result reduces to well known Kerr light deflection angle.

\section{General geodesic equations in KTN space time}
The KTN solution expressed in the Boyer-Lindquist coordinates $(ct, r, \vartheta, \varphi)$ given by the following metric [26, 27],

\begin{equation}
\fl
 ds^{2}=\frac{-\Delta}{\rho^{2}}[cdt+(2n\cos\vartheta-a\sin^{2}\vartheta)d\varphi]^{2}+\frac{\sin^{2}\vartheta}{\rho^{2}}[acdt-(r^{2}+n^{2}+a^{2})]^{2}+\frac{\rho^{2}}{\Delta}dr^{2}+\rho^{2}d\vartheta
\end{equation}
where, $\Delta=r^{2}-2mr+a^{2}-n^{2}$, $\rho^{2}=r^{2}+(n+a^{2}\cos\vartheta)^{2}$, $m=\frac{GM}{c^{2}}$ and $a=\frac{J}{cM}$ further c, G, M, J and $ n $ are the velocity of light in free space, gravitational constant, mass, angular momentum of the gravitating body and the NUT charge. Both $ m $ and $ n $ have the dimension of length. If we set $a=0$, the equation (1) will reduce to NUT solution.
\\Two horizons are located at the of roots of $r^{2}-2mr+a^{2}-n^{2}=0$ i.e
 \begin{equation}
 r_{\pm}=m\pm\sqrt{m^{2}-a^{2}+n^{2}}
 \end{equation}
Let us consider the light ray is moving through KTN space time, the related generalized geodesic equations can be obtained from the line element in equation (1) using Hamilton-Jacobi equation [28, 31] as:
\begin{equation}
  \rho^{2}\frac{\partial r}{\partial\tau}=\sqrt{R(r)}
 \end{equation}
 \begin{equation}
  \rho^{2}\frac{\partial \vartheta}{\partial\tau}=\sqrt{\Theta(\vartheta)}
 \end{equation}
 \begin{equation}
\rho^{2}\frac{\partial\varphi}{\partial\tau}=-[a-2n\cot\vartheta\csc\vartheta]E+L\csc^{2}\vartheta+\frac{a[E(r^{2}+n^{2}+a^{2})-aL]}{\Delta}
 \end{equation}
 \begin{eqnarray}
 \rho^{2}\frac{\partial(ct)}{\partial\tau}=-E[\sin^{2}\vartheta(a+2n\csc^{2}\vartheta)^{2}-4n\{n+a(1+\cos\vartheta)\}]\nonumber \\
 -(2n\cos\vartheta\csc^{2}\vartheta-a)L+\frac{(r^{2}+a^{2}+n^{2})}{\Delta}[E(r^{2}+a^{2}+n^{2})-aL]
 \end{eqnarray}
where,
$$
 R(r)=[E(r^{2}+a^{2})-aL]^{2}+n^{2}E[2a^{2}E-2aL+E(n^{2}+2r^{2})]-\Delta[K+(aE-L)^{2}]
$$
 and
$$
 \Theta(\vartheta)=K-\cos^{2}\vartheta[-a^{2}E^{2}+\frac{L^{2}}{\sin^{2}\vartheta}]+2n\cos\vartheta[2aE^{2}+2EL\csc^{2}\vartheta]-4n^{2}E^{2}\cot^{2}\vartheta
$$
The three constants of motion are $L$ (angular momentum of the particle along the direction of rotation), $E$ (energy of the particle) and $K$ (Carter constant).  For $K=0$ particle motion entirely remains in the equatorial plane.
\section{Equatorial geodesic equations in KTN space time}
The main objective of this paper is to calculate the light deflection angle for KTN geometry in equatorial plane. So at this point we will convert the generalized geodesic equations to equatorial geodesic equations. The conditions for equatorial plane are $\vartheta=\frac{\pi}{2}$ and $ K=0$. Using these conditions, the modified equatorial geodesic equations can be obtained.
\\The modified version of equation (3) will be,
\begin{equation}
 (r^{2}+n^{2})^{2}\dot{r}^{2}=[E(r^{2}+a^{2})-aL]^{2}+n^{2}E[2a^{2}E-2aL+E(n^{2}+2r^{2})]-\Delta(aE-L)^{2}
\end{equation}
Here ($^{.}$) indicates the derivative with respect to affine parameter $\tau$. Let us use the value of $\Delta=a^{2}-2mr+a^{2}-n^{2}$ in the above equation, we have,
\begin{equation}
\fl
 (r^{2}+n^{2})^{2}\dot{r}^{2}=[E(r^{2}+a^{2})-aL]^{2}+n^{2}E[2a^{2}E-2aL+E(n^{2}+2r^{2})]-(a^{2}-2mr+a^{2}-n^{2})(aE-L)^{2}
\end{equation}
After some simplification,
\begin{eqnarray}
(r^{2}+n^{2})^{2}\dot{r}^{2}=E^{2}(r^{4}+n^{4})+(L-aE)^{2}(2mr+n^{2})-r^{2}(L^{2}-a^{2}E^{2})\nonumber \\
+n^{2}[2a^{2}E^{^{2}}-2aEL+2r^{2}E^{2}]
\end{eqnarray}
As we have already shown in our previous work [12], it is possible to show that impact parameter is the ratio of $L$ and $E$ [31, page 328]. Following [5, equation no. 7], we can write the impact parameter $b_{s}\equiv sb=s(\frac{L}{E})$, where $s=+1$ for prograde and $s=-1$ for retrograde orbit of light ray and $b$ is the positive magnitude of the impact parameter.So the new form of the above equation is,
\begin{eqnarray}
 (r^{2}+n^{2})^{2}\dot{r}^{2}=L^{2}[\frac{r^{4}+n^{4}}{b^{2}}+(1-\frac{a}{b_{s}})^{2}(2mr+n^{2})-r^{2}(1-\frac{a^{2}}{b^{2}})\nonumber \\
 +2n^{2}\{(1-\frac{a}{b_{s}})-(1-\frac{a^{2}}{b^{2}})+\frac{2r^{2}}{b^{2}}\}]
\end{eqnarray}
or,
\begin{eqnarray}
 (r^{2}+n^{2})^{2}\dot{r}^{2}=L^{2}[\frac{r^{4}}{b^{2}}+(1-\frac{a}{b_{s}})^{2}(2mr+n^{2})-r^{2}(1-\frac{a^{2}}{b^{2}})\nonumber \\
 +2n^{2}\{(1-\frac{a}{b_{s}})-(1-\frac{a^{2}}{b^{2}})\}+\frac{n^{2}}{b^{2}}(4r^{2}+n^{2})]
\end{eqnarray}
As r obtains a local extremum for the closest approach $r_{0}$, we can write:
 $$\dot{r}|_{r=r_{0}}=0$$
 Thus, from equation (13),
\begin{eqnarray}
 \frac{r_{0}^{4}}{b^{2}}+(1-\frac{a}{b_{s}})^{2}(2mr_{0}+n^{2})-r_{0}^{2}(1-\frac{a^{2}}{b^{2}})\nonumber \\
 +2n^{2}\{(1-\frac{a}{b_{s}})-(1-\frac{a^{2}}{b^{2}})\}+\frac{n^{2}}{b^{2}}(4r_{0}^{2}+n^{2})=0
\end{eqnarray}
or,
\begin{eqnarray}
 \frac{r_{0}^{2}}{b^{2}}=-(1-\frac{a}{b_{s}})^{2}(2\frac{m}{r_{0}}+\frac{n^{2}}{r^{2}_{0}})+(1-\frac{a^{2}}{b^{2}})\nonumber \\
 -2\frac{n^{2}}{r^{2}_{0}}\{(1-\frac{a}{b_{s}})-(1-\frac{a^{2}}{b^{2}})\}-\frac{n^{2}}{b^{2}}(4+\frac{n^{2}}{r^{2}_{0}})
\end{eqnarray}
From equation (4), it is clear that,
 \begin{equation}
 \dot{\vartheta}|_{\vartheta=\frac{\pi}{2}}=0
 \end{equation}
From equation (5) we can have,
 \begin{equation}
 (r^{2}+n^{2})^{2}\dot{\varphi}=-(aE-L)+\frac{a}{\Delta}[E(r^{2}+a^{2}+n^{2})-aL]
 \end{equation}
 After some simplification we can write,
 \begin{equation}
 \dot{\varphi}=\frac{L}{\Delta}[\frac{2mra}{b}+r^{2}-2mr+n^{2}(\frac{2a}{b}-1)]
 \end{equation}
From equation (6) we have,
 \begin{equation}
 \fl
 (r^{2}+n^{2})^{2}c\dot{t}=-E[(a+2n)^{2}-4n(n+a)]
 +aL+\frac{(r^{2}+a^{2}+n^{2})}{\Delta}[E(r^{2}+a^{2}+n^{2})-aL]
 \end{equation}
 After some simplification,
 \begin{equation}
 (r^{2}+n^{2})^{2}c\dot{t}=\frac{1}{\Delta}[-E\{a^{2}(a^{2}-2mr+m^{2}-n^{2})-(r^{2}+a^{2}+n^{2})^{2}\}-2aL(mr+n^{2})]
 \end{equation}
 Equation (11), (16) and (18) are the equatorial geodesic equations governing the motion of light ray under KTN space time geometry.

\section{Radius of Photon Sphere}
At this part of the paper, we will obtain the equatorial circular orbit (representing photon sphere) with radius $r_{ph}$. The photon radius $r_{ph}$ is defined by the condition $R(r) = \frac{dR(r)}{dr} = 0$, for $r=r_{ph}$ [12, 13]. So, from equation (9) we can have the following equations,
$$E^{2}(r^{4}+n^{4})+(L-aE)^{2}(2mr+n^{2})-r^{2}(L^{2}-a^{2}E^{2})+n^{2}[2a^{2}E^{^{2}}-2aEL+2r^{2}E^{2}]=0$$
or,
$$E^{2}+(\frac{2m}{r_{ph}^{3}}+\frac{n^{2}}{r_{ph}^{4}})(L-aE)^{2}-\frac{1}{r_{ph}^{2}}(L^{2}-a^{2}E^{2})+E^{2}(\frac{n^{4}}{r_{ph}^{4}})$$
$$+\frac{n^{2}}{r_{ph}^{4}}[2a^{2}E^{2}-2aEL]+\frac{2n^{2}E^{2}}{r_{ph}^{2}}=0$$

and
$$\frac{d}{dr}[E^{2}+(\frac{2m}{r_{ph}^{3}}+\frac{n^{2}}{r_{ph}^{4}})(L-aE)^{2}-\frac{1}{r_{ph}^{2}}(L^{2}-a^{2}E^{2})+E^{2}(\frac{n^{4}}{r_{ph}^{4}})$$
$$+\frac{n^{2}}{r_{ph}^{4}}[2a^{2}E^{2}-2aEL]+\frac{2n^{2}E^{2}}{r_{ph}^{2}}]=0$$
or,
$$(-\frac{3m}{r_{ph}^{4}}-\frac{2n^{2}}{r_{ph}^{5}})(L-aE)^{2}+\frac{1}{r_{ph}^{3}}(L^{2}-a^{2}E^{2})-\frac{2E^{2}n^{4}}{r_{ph}^{5}}$$
$$-\frac{2n^{2}}{r_{ph}^{5}}(2a^{2}E^{2}-2aEL)-\frac{2n^{2}E^{2}}{r_{ph}^{3}}=0$$
or,
$$r_{ph}^{2}\{(L^{2}-a^{2}E^{2})-2n^{2}E^{2}\}-r_{ph}\{3m(L-aE)^{2}\}-\{2n^{2}(L-aE)^{2}+2E^{2}n^{4}+2n^{2}(2a^{2}E^{2}-2aEL)\}=0$$
solving the above equation for $r_{ph}$ we have,
$$r_{ph}=[3m(L-aE)^{2}\pm [9m^{2}(L-aE)^{2}-4n^{2}\{(L^{2}-a^{2}E^{2})-2n^{2}E^{2}\}\{2(L-aE)^{2}$$
$$+2E^{2}(n^{2}+a^{2})-2aEL\}]^{\frac{1}{2}}]/[2\{(L^{2}-a^{2}E^{2})-2n^{2}E^{2}\}]$$
The above equation represents the equatorial circular orbit. For $n=0$, we have,
$$r_{ph}=3m\frac{L-aE}{L+aE}$$
This is the expression of radius of equatorial circular orbit [31, page 329, equation (80)].

\section{Equatorial light deflection angle}
The light deflection angle can be expressed as [32, page: 188],
\begin{equation}
 \alpha=2\int_{r_{0}}^{\infty}(\frac{d\varphi}{dr}).dr -\pi
\end{equation}
Now using equations (11) and (16) in the equation (19) we can have,
\begin{eqnarray}
 \fl
\alpha=2\int_{r_{0}}^{\infty}[\frac{2mra}{b}+r^{2}-2mr+n^{2}(\frac{2a}{b}-1)]/\Delta[\frac{r^{4}}{b^{2}}+(1-\frac{a}{b_{s}})^{2}(2mr+n^{2})\nonumber \\
-r^{2}(1-\frac{a^{2}}{b^{2}})+2n^{2}\{(1-\frac{a}{b_{s}})-(1-\frac{a^{2}}{b^{2}})\}+\frac{n^{2}}{b^{2}}(4r^{2}+n^{2})]^{\frac{1}{2}} -\pi
\end{eqnarray}
or,
\begin{eqnarray}
 \fl
\alpha=2\int_{r_{0}}^{\infty}[1-\frac{2m}{r}+\frac{2ma}{br}+\frac{n^{2}}{r^{}}(\frac{2a}{b}-1)]/[1-\frac{2m}{r}+\frac{a^{2}}{r^{2}}-\frac{n^{2}}{r^{2}}][\frac{r^{4}}{b^{2}}+(1-\frac{a}{b_{s}})^{2}(2mr+n^{2})\nonumber \\
-r^{2}(1-\frac{a^{2}}{b^{2}})+2n^{2}\{(1-\frac{a}{b_{s}})-(1-\frac{a^{2}}{b^{2}})\}+\frac{n^{2}}{b^{2}}(4r^{2}+n^{2})]^{\frac{1}{2}}-\pi
\end{eqnarray}
let us introduce a new variable $x=\frac{r_{0}}{r}$. So,
$$dx=-\frac{r_{0}dr}{r^{2}}$$
or,
$$\frac{dx}{r_{0}}=-\frac{dr}{r^{2}}$$
the limits will change now as and when $r\longrightarrow \infty$, then $x\longrightarrow 0$ and when $r\longrightarrow r_{0}$, then $x\longrightarrow 1$. Using this in above equation we have,
\begin{equation}
 \alpha=2\int_{0}^{1}\frac{f_{1}}{f_{2}\sqrt{f_{3}}}.dx-\pi
\end{equation}
where,
$$f_{1}=1-2hxF-l^{2}x^{2}(2F-1)$$
$$f_{2}=1-2hx+\hat{a}^{2}h^{2}x^{2}-l^{2}x^{2}$$
and
$$f_{3}=\frac{r_{0}^{2}}{b^{2}}+F^{2}x^{2}(2hx+l^{2}x^{2})-Gx^{2}+2l^{2}x^{4}(F-G)+\hat{l}^{2}x^{2}(4+l^{2}x^{2})$$
further, $h=\frac{m}{r_{0}}$ and $l^{2}=\frac{n^{2}}{r_{0}^{2}}$, $\hat{l}^{2}=\frac{n^{2}}{b^{2}}$ and $\hat{a}=\frac{a}{m}$. So the mass and NUT factor are now represent by $h$ and $l$. We again followed [8, 9] and substituted $G=1-(\frac{a}{b})^{2}=1-\hat{a}^{2}(\frac{m}{b})^{2}$ and $F=1-(\frac{a}{b_{s}})=1-s\hat{a}\frac{m}{b}$. Thus for zero rotation $(\hat{a}=0)$, $F=G=1$.
\\Now we will use the expression $\frac{r_{0}^{2}}{b^{2}}=-(1-\frac{a}{b_{s}})^{2}(2\frac{m}{r_{0}}+\frac{n^{2}}{r^{2}_{0}})+(1-\frac{a^{2}}{b^{2}})
 -2\frac{n^{2}}{r^{2}_{0}}\{(1-\frac{a}{b_{s}})-(1-\frac{a^{2}}{b^{2}})\}-\frac{n^{2}}{b^{2}}(4+\frac{n^{2}}{r^{2}_{0}})=G-F^{2}(2h+l^{2})-2l^{2}(F-G)-\hat{l}^{2}(4+l^{2})$ from equation (13) in $f_{3}$ and rearranging we get,
 $$f_{3}=(G-4\hat{l}^{2})(1-x^{2})-2hF^{2}(1-x^{3})-[F^{2}l^{2}+2l^{2}(F-G)+\hat{l}^{2}l^{2}](1-x^{4})$$
Let us substitute $G-4\hat{l}^{2}=A$, so the new form of $f_{3}$ will be,
$$f_{3}=A(1-x^{2})-2hF^{2}(1-x^{3})-[F^{2}l^{2}+2l^{2}(F-G)+\hat{l}^{2}l^{2}](1-x^{4})$$
or,
$$f_{3}=A(1-x^{2})[1-\frac{2hF^{2}}{A}(\frac{1-x^{3}}{1-x^{2}})-\frac{\{F^{2}l^{2}+2l^{2}(F-G)+\hat{l}^{2}l^{2}\}}{A}(1+x^{2})]$$
or,
$$f_{3}=A(1-x^{2})[1-\frac{2hF^{2}}{A}(\frac{1-x^{3}}{1-x^{2}})][1-\frac{\{F^{2}l^{2}+2l^{2}(F-G)+\hat{l}^{2}l^{2}\}}{A}(1+x^{2})\{1-\frac{2hF^{2}}{A}(\frac{1-x^{3}}{1-x^{2}})\}^{-1}]$$  following [10], let us substitute,  $\frac{F^{2}h(1-x^{3})}{A(1-x^{2})}=\delta_{2}$ and $\frac{\{F^{2}l^{2}+2l^{2}(F-G)\}(1+x^{2})}{A}=\delta_{1}$. So the new form of $f_{3}$ using $\delta_{2}$ and $\delta_{1}$ is,
$$f_{3}=A(1-x^{2})[1-2\delta_{2}][1+(\delta_{1}+\frac{l^{2}\hat{l}^{2}(1+x^{2})}{A})(1-2\delta_{2})]^{-1}$$
Putting this values of $f_{1}$, $f_{2}$ and $f_{3}$ in equation (22), we get,
\begin{equation}
 \alpha=2\int_{0}^{1}\frac{f_{1}f_{2}^{-1}}{\sqrt{A}\sqrt{1-x^{2}}\sqrt{1-2\delta_{2}}\sqrt{1+(\delta_{1}+\frac{l^{2}\hat{l}^{2}(1+x^{2})}{A})(1-2\delta_{2})^{-1}}}.dx-\pi
 \end{equation}
 Now rearranging the above equation we can write.
\begin{equation}
 \alpha=2\int_{0}^{1}\frac{dx}{\sqrt{A}\sqrt{1-x^{2}}}f_{1}f_{2}^{-1}(1-2\delta_{2})^{-\frac{1}{2}}[1+(\delta_{1}+\frac{l^{2}\hat{l}^{2}(1+x^{2})}{A})(1-2\delta_{2})^{-1}]^{-\frac{1}{2}}-\pi
 \end{equation}
For the weak deflection limit, following [8, 9, 10] one can assume, $n,m\ll r_{0}$, in other words, $h,l\ll 1$. So the equation (24) can be expanded in the Taylor series in terms of both $h$ and $l$. Here we calculate the deflection angle, considering contribution up to fourth order terms in mass and NUT factor only and write,
$$\alpha=2\int_{0}^{1}\frac{dx}{\sqrt{A}\sqrt{1-x^{2}}}f_{1}[1+2hx+x^{2}h^{2}(4-\hat{a}^{2})+x^{3}h^{3}(8-4\hat{a}^{2})+x^{4}h^{4}(\hat{a}^{4}-12\hat{a}^{2}+16)]$$
$$[1+x^{2}l^{2}(1+2hx+x^{2}h^{2}\{4-\hat{a}^{2}\})+x^{4}l^{4}][1+\delta_{2}+\frac{3}{2}\delta_{2}^{2}+\frac{5}{2}\delta_{2}^{3}+\frac{35}{8}\delta_{2}^{4}]$$
$$[1+\delta_{1}(1+2\delta_{2}+4\delta_{2}^{2})+\frac{3}{8}\delta_{1}^{2}+\frac{l^{2}\hat{l}^{2}(1+x^{2})}{2A}]-\pi$$

Let us substitute $s_{0}=-\hat{a}^{2}+4-4F$, $s_{1}=-4\hat{a}^{2}+8+2F\hat{a}^{2}-8F$ and $s_{2}=\hat{a}^{4}-12\hat{a}^{2}+16+8\hat{a}^{2}F-16F$ by following [12]. Then multiplying term by term and retaining only up to fourth order in both $h$ and $l, \hat{l}$ we will get form the above equation,
$$\alpha=2\int_{0}^{1}\frac{dx}{\sqrt{A}\sqrt{1-x^{2}}}[1+h\{2x(1-F)+6l^{2}x^{3}(1-F)+\frac{\delta_{2}}{h}+\frac{2l^{2}x^{2}(1-F)\delta_{2}}{h}+x(1-F)\delta_{1}+\frac{3\delta_{1}\delta_{2}}{2h}\}$$
$$+h^{2}\{x^{2}s_{0}+l^{2}x^{4}(16-16F-3\hat{a}^{2}+2\hat{a}^{2}F)+\frac{2x(1-F)\delta_{2}}{h}+\frac{6l^{2}x^{3}\delta_{2}(1-F)}{h}+\frac{3}{2h^{2}}\delta^{2}_{2}+\frac{3l^{2}x^{2}(1-F)\delta^{2}_{2}}{h^{2}}$$
$$\frac{x^{2}s_{0}\delta_{1}}{2}+\frac{15\delta_{1}\delta^{2}_{2}}{4h^{2}}+\frac{3\delta_{1}\delta_{2}(1-F)x}{h}\}+h^{3}\{x^{3}s_{1}+\frac{x^{2}s_{0}\delta_{2}}{h}+\frac{3x(1-F)\delta_{2}^{2}}{h^{2}}+\frac{5\delta_{2}^{3}}{2h^{3}}\}+h^{4}\{x^{4}s_{2}+\frac{x^{3}\delta_{2}s_{1}}{h}$$
$$+\frac{3}{2}x^{2}s_{0}\frac{\delta_{2}^{2}}{h^{2}}+5x(1-F)\frac{\delta_{2}^{3}}{h^{3}}+\frac{35}{8}\frac{\delta_{2}^{4}}{h^{4}}\}+l^{2}\{2x^{2}(1-F)-\frac{\delta_{1}}{2l^{2}}+\frac{\hat{l}^{2}(1+x^{2})}{2A}\}$$
$$+l^{4}\{2x^{4}(1-F)+\frac{x^{2}\delta_{1}}{2n^{2}}(1-F)+\frac{3\delta_{2}^{2}}{8n^{4}}\}]-\pi$$
Now integrating the above equation term by term we get,

\begin{eqnarray}
\fl
\alpha=c_{0}\pi+4h[c_{1}-\frac{l^{2}(1-F)}{\sqrt{A}}\{2+(\frac{8}{3}-\frac{\pi}{2})\frac{F}{A}+\frac{5}{4A}(F^{2}+2(F-G))\}\nonumber \\
\fl
+(\frac{7}{2}-\frac{3\pi}{8})\frac{l^{2}F^{2}(F^{2}+2(F-G))}{A^{\frac{5}{2}}}]+h^{2}[-4c_{2}+\frac{15\pi}{4}d_{2}
+\frac{l^{2}}{\sqrt{A}}\{\frac{3\pi}{8}(16-16F-3\hat{a}^{2}+2\hat{a}^{2}F)\nonumber \\
\fl
+(-24+\frac{45\pi}{4})\frac{F^{2}(1-F)}{A}+\frac{F^{4}(1-F)}{A^{2}}(\frac{105\pi}{8}-32)+\frac{7\pi s_{0}}{16A}(F^{2}+2(F-G))\nonumber \\
\fl
+(\frac{825\pi}{32}-50)\frac{F^{4}(F^{2}+2(F-G))}{A^{3}}+(\frac{81\pi}{8}-18)\frac{F^{2}(1-F)(F^{2}+2(F-G))}{A^{3}}\}]\nonumber \\
\fl
+h^{3}\Big(\frac{122}{3}c_{3}-\frac{15\pi}{2}d_{3}\Big)+h^{4}\Big(-130c_{4}+\frac{3465\pi}{64}d_{4}\Big)
+l^{2}[\frac{(1-F)\pi}{\sqrt{A}}+\frac{3\pi\{(F^{2}+2(F-G))+\hat{l}^{2}\}}{4A^{\frac{3}{2}}}]\nonumber\\
\fl
+l^{4}[\frac{(1-F)\pi}{4\sqrt{A}}+\frac{7\pi(1-F)}{8A^{\frac{3}{2}}}(F^{2}+2(F-G))+\frac{57\pi(F^{2}+2(F-G))^{2}}{64A^{\frac{5}{2}}}]
\end{eqnarray}
where,
\numparts
\begin{equation}
c_{0}=\frac{1}{\sqrt{A}}-1
\end{equation}
\begin{equation}
c_{1}=\frac{F^{2}+A-FA}{A^{\frac{3}{2}}}
\end{equation}
\begin{equation}
c_{2}=\frac{F^{2}}{A}c_{1}
\end{equation}
\begin{equation}
d_{2}=\frac{1}{15A^{\frac{5}{2}}}[15F^{4}-4A(F-1)(3F^{2}+2A)-2A^{2}\hat{a}^{2}]
\end{equation}
\begin{eqnarray}
c_{3}=\frac{1}{61A^{\frac{7}{2}}}[61F^{6}-A(F-1)(45F^{4}+32F^{2}A+16A^{2})\nonumber\\
-4G^{2}\hat{a}^{2}(2F^{2}+2A-FA)]
\end{eqnarray}
\begin{equation}
d_{3}=\frac{F^{2}}{A}d_{2}
\end{equation}
\begin{equation}
c_{4}=\frac{F^{2}}{65A^{\frac{9}{2}}}[65F^{6}-49(F-1)F^{4}A+8F^{2}A^{2}s_{0}+2s_{1}A^{3}]
\end{equation}
\begin{eqnarray}
d_{4}=\frac{1}{1155A^{\frac{9}{2}}}[1155F^{8}-840(F-1)F^{6}A+140F^{4}s_{0}A^{2}\nonumber\\
+40s_{1}F^{2}A^{3}+8s_{2}A^{4}]
\end{eqnarray}
\endnumparts
The above substitution were followed from [8, 9, 12].
\\The above equation (25) represents the equatorial deflection of light under KTN geometry in weak field limit. This expression is a function of mass, rotation and NUT factor. As mentioned earlier in text after equation (25), the assumption used for derivation here is that both $n$ and $m\ll r_{0}$. Thus this result will be reduced to that of Kerr metric by setting $n=0$.
\\If we set NUT factor equal to zero in the equation (25), we will get the expression of deflection for light by Kerr mass obtained by Aazami et al. [9, equation no. (B17)].
\\If we set mass and rotation equal to zero i.e $h=0$, $F=G=1$ and $A=\sqrt{1-4\hat{l}^{2}}$ in the equation (25), we will get,
\begin{equation}
 \alpha=l^{2}[\frac{3\pi}{4}\frac{[1+\hat{l}^{2}]}{\sqrt{1-4\hat{l}^{2}}}]+l^{4}[\frac{57\pi}{64\sqrt{1-4\hat{l}^{2}}}]
\end{equation}
This is the amount of deflection of light ray occurring only due to NUT factor. A hypothetical massless, static body with non zero NUT factor can influence the curvature of space time.
\section{Discussion of results}
We have obtained the expression of event horizon from equation (2) which is $r_{\pm}=m\pm\sqrt{m^{2}-a^{2}+n^{2}}$. The above expression represents two event horizons and for $a^{2}>m^{2}+n^{2}$, there will be no event horizon and a naked singularity will appear, which violates the causality of the space time and is forbidden according to the Penrose's cosmic censorship conjecture [28]. It clearly gives a limit to the value of ($m^{2}+n^{2}$) i.e it must be greater than $a^{2}$.
\\To understand the physical significance of the calculation done in this paper, we plot bending angle $(\alpha)$ against various physical parameters ($\hat{a}$, $l$, $b$) in Fig1, Fig2, Fig3 respectively by taking Sun as a test case. In our previous work [12], we obtained the equatorial deflection angle for Kerr-Newman (rotation with electric charge) body. In [12], we used Sun as a test case and plotted bending angle against various physical parameters by considering that Sun has some static charge and the closest approach of the light ray was the radius of the Sun. Following [12], here we consider that Sun has some NUT charge $l=1.413850947\times10^{-6}$ and the closest approach is the radius of Sun $(6.955\times10^{8}m)$. The value of $l=n/r_{0}$ is chosen in such way that $(m^{2}+n^{2})$ is always greater than $a^{2}$.
\\Fig1 clearly shows the difference between prograde and retrograde motion with respect to zero rotation Taub-NUT case. The nature of bending angle versus rotation parameter curve is similar to the result obtained by Iyer and Hansen [5, 6] and Chakraborty and Sen [12] for Kerr and Kerr-Newman equatorial bending.
\\From the Fig2, we can say, deflection of the light ray increases with the increase of NUT factor of the body. When NUT factor is zero, light ray gets the minimum deflection. We know that the presence of static charge reduces the amount of deflection of light [12, Fig2]. So we can see that NUT factor and static charge influence the space time geometry in opposite direction. The presence of NUT factor increases the light deflection angle compared to zero field Kerr case. On the other hand, the presence of static charge decreases the light deflection angle with respect to Kerr case.
\\Fig3 shows the change of bending angle with impact parameter. Though the pattern is similar as given by Iyer and Hansen [5, 6] and Chakraborty and Sen [12], but the prograde, retrograde and zero spin Taub-NUT plot overlap with each other as the difference between them is small for Sun. So following [12], we consider a slow rotating pulsar PSR J 1748-2446 [33] as a test case and plot the bending angle versus impact parameter curve in Fig4. We consider the pulsar has some NUT charge $l=1.413850947\times10^{-6}$ and closest approach is the radius of the pulsar i.e 20km. The pattern of the graph is similar as given by Iyer and Hansen [5, 6] and Chakraborty and Sen [12]. For this pulsar we take $m=1.99$km and calculate $a$ as 0.96km from the input values of time period $T=1.393$ms and $r_{0}=20$km as listed by Nu$\tilde{n}$ez and Nowakowski [33].
\section{Conclusions}
From the above study we may conclude the following:\\
1. Expression for equatorial deflection of light due to a KTN body has been calculated considering contributions from mass and NUT factor up to fourth order terms.\\
2. NUT factor has a noticeable effect on the path of the light ray. When compared with Kerr expression for bending, we find that there are some extra terms in the expression for deflection which occur due to the presence of NUT factor. If we set the NUT factor equal to zero, deflection angle will reduce to that of Kerr deflection angle. When we compare the effect of NUT factor with that of static charge, we can see that these two parameters have opposite effect on the space-time geometry.\\
3. As we know, with KTN metric even with a hypothetical body of zero mass and non-zero NUT factor, one can have some effect on the space-time curvature.

\begin{figure}
 \includegraphics[width=12cm]{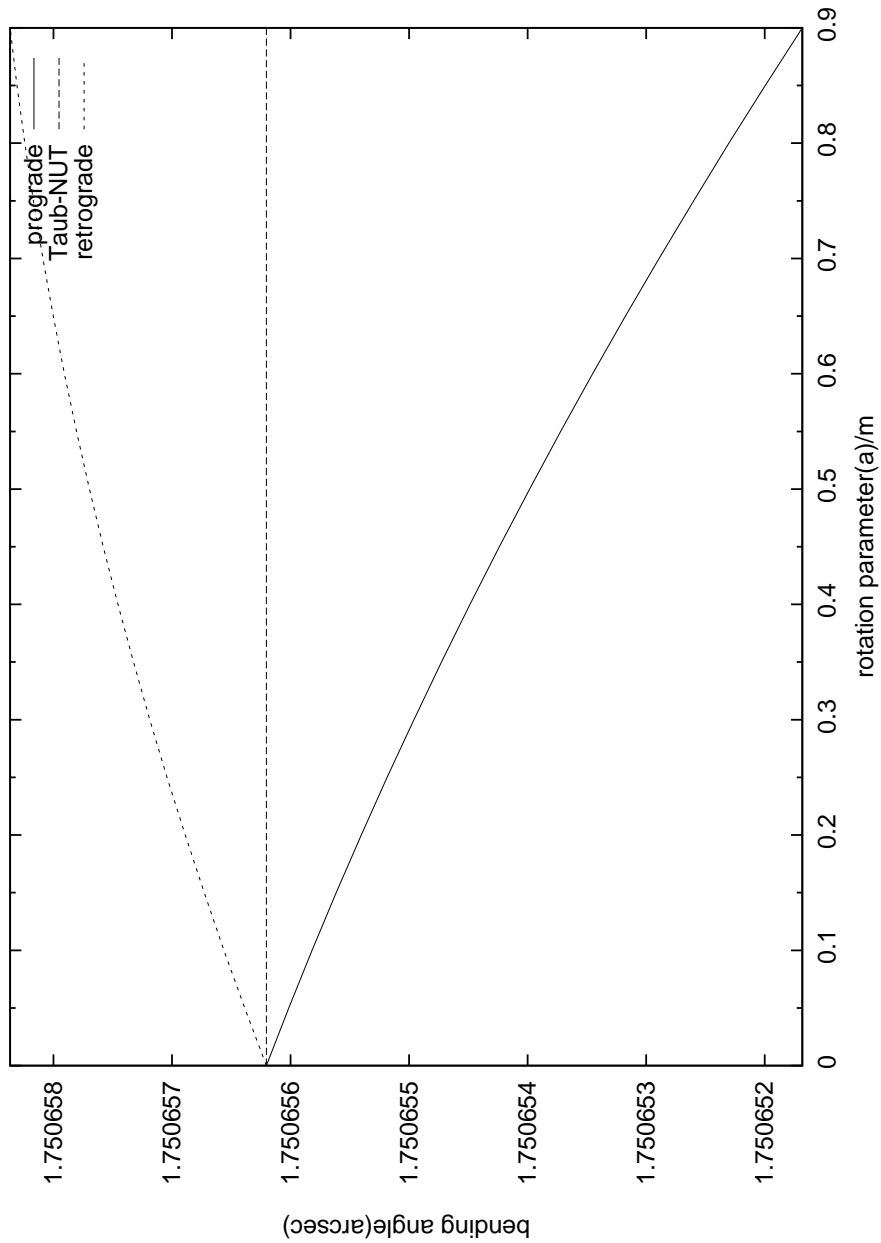}%
 \caption{\label{}Bending angle (arcsec) as a function of rotation parameter (a/m) with constant NUT factor $l=1.413850947\times10^{-6}$ and impact parameter is one solar radius. Here three different curves represent the prograde (when the light ray moves in the direction of rotation of the body), corresponding Taub-NUT (zero rotation) and retrograde (when the light ray moves in the opposite direction of the rotation of the body) motion of light ray.}
 \end{figure}

 \begin{figure}
\includegraphics[width=12cm]{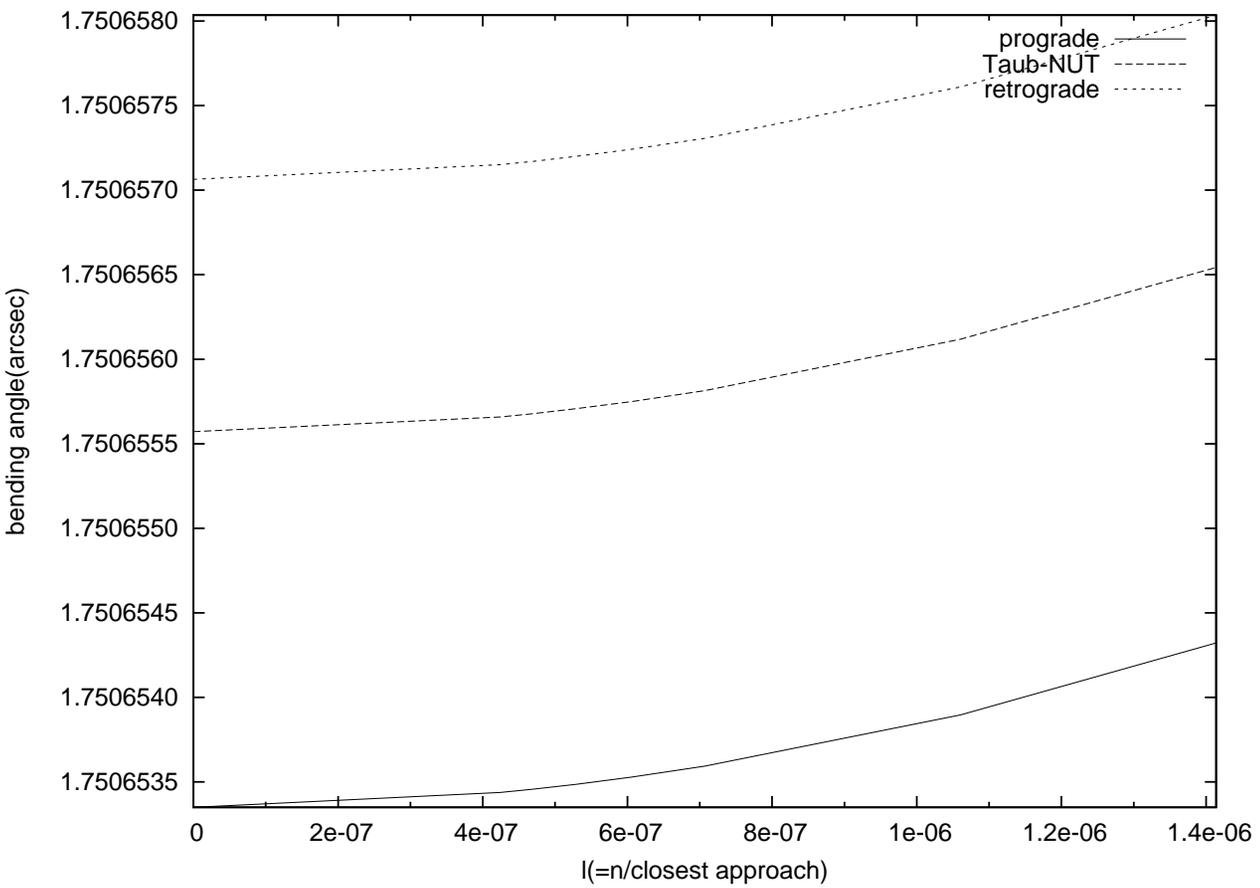}%
\caption{\label{} Bending angle (arcsec) as a function of charge $l=n/r_{o}$ with constant rotation parameter $\hat{a}=0.5$ and impact parameter is one solar radius. Geometries are explained in the caption for Fig.1.}
\end{figure}

\begin{figure}
\includegraphics[width=12cm]{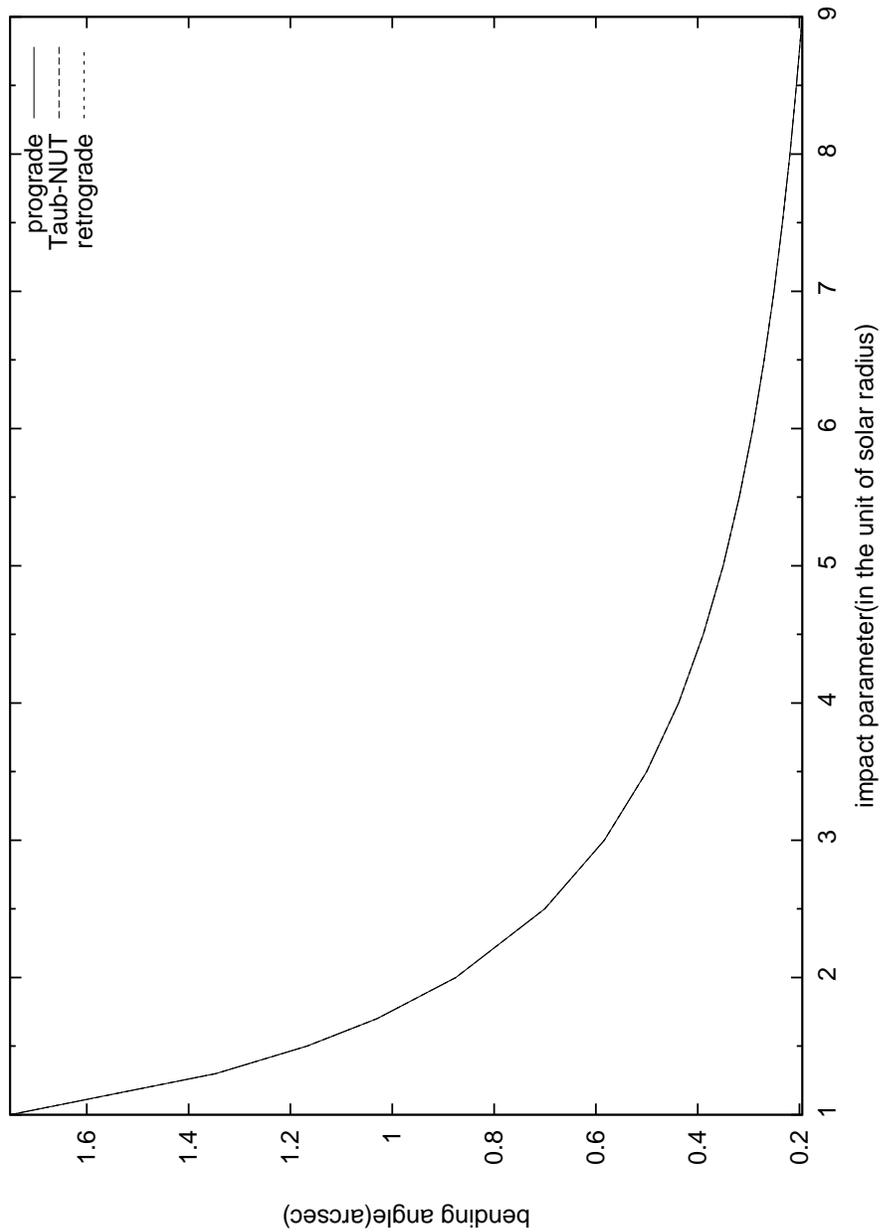}%
\caption{\label{} Bending angle (arcsec) as a function of impact parameter in the unit of solar radius with constant rotation ($\hat{a}=0.5$) and NUT factor $l=1.413850947\times10^{-6}$. Geometries are explained in the caption for Fig.1. All the plots for prograde, Taub-NUT and retrograde merge together for Sun.}
\end{figure}

\begin{figure}
\includegraphics[width=12cm]{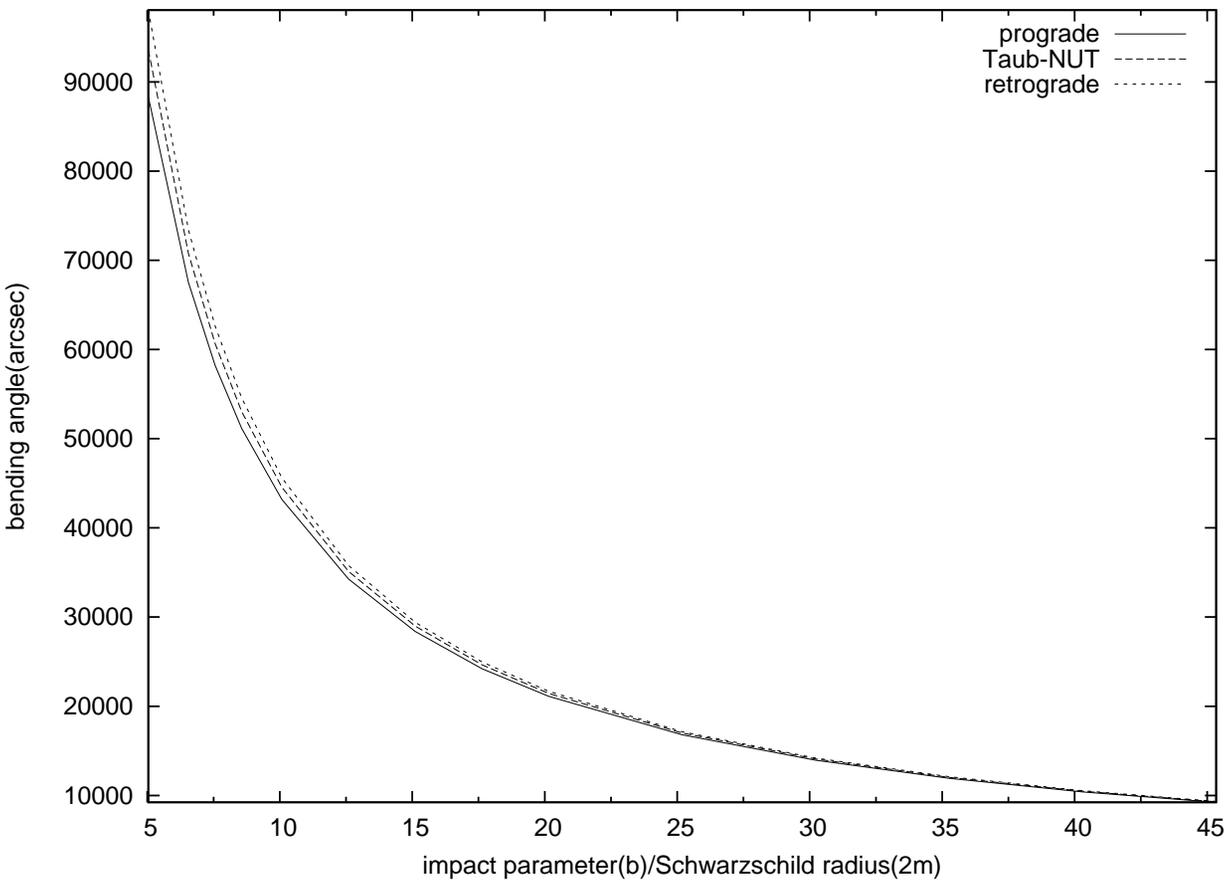}%
\caption{\label{} Bending angle (arcsec) as a function of impact parameter $b/r_{g}$ with constant rotation ($\hat{a}=0.5$) and NUT factor $l=1.413850947\times10^{-6}$. Geometries are explained in the caption for Fig.1.}
\end{figure}

\end{document}